\newcommand {\fd} {\rightarrow}
\newcommand {\be} {\begin{equation}}
\newcommand {\ee} {\end  {equation}}
\newcommand {\ba} {\begin{eqnarray}}
\newcommand {\ea} {\end  {eqnarray}}
\newcommand {\bay} {\begin{array}}
\newcommand {\eay} {\end {array}}

\documentclass[10pt]{article}

\usepackage{graphicx}

\begin{document}

\title{Thermodynamics of a Tiling Model}
\author{Luca Leuzzi$^{\star}$ and Giorgio Parisi$^{\star\star}$}

\maketitle
\begin{center}
{\small {$\star$ Instituut voor  Theoretische Fysica, 
Universiteit van Amsterdam}}\\
{\small {Valckenierstraat 65, 1018 XE Amsterdam, The Netherlands }}

{\small {$\star$$\star$  Dipartimento di Fisica and Infn, Universit\`a 
di Roma ``La Sapienza''}}\\
{\small{P.A.Moro 2, 00185 Roma, Italy}}
\end{center}

\begin{abstract}
A particular, two-dimensional, tiling model, composed by the so-called Wang 
tiles \cite{gruenbaum} has been studied at finite temperature by 
Monte Carlo numerical simulations.
In absence of any thermal bath the Wang tiles give the opportunity of
 building a very large number of non-periodic tilings.
We can construct a local Hamiltonian such that only perfectly matched tilings
 are ground states with zero energy.
This Hamiltonian has a very large degeneracy.
The  thermodynamic behaviour of such a system seems to show a continuous
 phase transition at non zero 
temperature. An order parameter with non-trivial features is proposed. 
Under the critical temperature the model exhibits aging properties. The  
fluctuation-dissipation theorem is violated.
\end{abstract}

\section{Introduction to the model}
 In this work we propose a model that 
shows  a non-trivial thermodynamic behaviour 
due to  its particular geometric structure.

\begin{figure}[!htb]
\begin{center}
\label{fig:wangtiles}
	\includegraphics[width=5 cm,height= 5 cm]{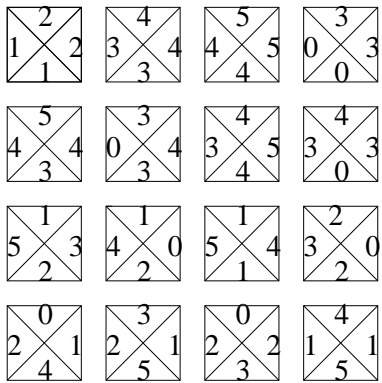}
\caption{\small{The sixteen Wang tiles. The six 
different types of edges, or 'colours'
are indicated  here by numbers ($0$, $1$, $2$, $3$, $4$, $5$). 
The tiles shown  are the minimum set  allowing
aperiodic tiling}}
\end{center}
\end{figure}

This two-dimensional model is built using  square tiles, called Wang's tiles,
 on a square lattice. The edges of these tiles can be of six different 
'colours'. But of all the possible types ($6^4$)
that can be created changing the 
colours  just a particular group of sixteen, found by Ammann \cite{gruenbaum},
 is taken into account (see figure \ref{fig:wangtiles}). 

This is one of the minimal set of Wang tiles 
such that the corresponding tilings exist and are non-periodic tilings.
A tiling is a configuration of tiles placed edge-to-edge on the plane, where
all contiguous edges  have the same colour. 
If at least one   tiling  is allowed and none of the possible tilings shows a 
 periodic pattern, then the set of tiles composing them is
 called {\it aperiodic}. Other aperiodic sets of 
tiles are often used as models for 
quasi-crystal
 materials \cite{quasi-crist},
 but this is not the present case.

The Wang tiles were the first aperiodic tiles  to be discovered, in 1966 by 
Berger \cite {berger}.
They were initially important, and they still are, 
 because of their use in problems of 
mathematical logic \cite{gruenbaum}\cite{Culik/Kari}.
The use that we make of them in this work is, anyway,  far away
 from that point of view. Indeed we look at the behaviour of a system built
by Wang tiles in a thermal bath, defining thermodynamic observables
on these tilings.
The main stimulus  to study this model has been 
  the high degeneracy of the
perfectly matched configurations and their non-trivial aperiodic structure.

Putting the system in a thermal bath  allows 
for translations of the tiles also in positions where 
there is no matching between edges of neighbours
tiles, forming in this way a unmatched-tiling.
 We can assign an energy to each configuration which is equal to 
the number of links such that the colours of the facing edges are 
different.
The energy of the 
exactly matched  configurations, that from now on we will call
 ground states, is then zero.

If we  label the  type of tile in the position given
  by the coordinates $(x,y)$
in the plane by $T_{x,y}$ and the type of its four edges respectively 
towards South, East, North and West by
$T^{(S)}_{x,y}$, $T^{(E)}_{x,y}$, $T^{(N)}_{x,y}$ and $T^{(W)}_{x,y}$
we can write an Hamiltonian for this system in the following way:
\be
{\cal{H}}=\hspace{-1 mm}\sum^{1,L-1}_{(x,y)}\hspace{-1 mm}
  \left(1-\delta\left(T^{(E)}_{x,y}-T^{(W)}_{x+1,y}\right)\right) + 
 \hspace{-1 mm} \sum^{1,L-1}_{(x,y)}\hspace{-1 mm}  
\left(1-\delta\left(T^{(N)}_{x,y}-T^{(S)}_{x,y+1}\right)\right).
\label{hamil}
\ee
\noindent $L$ is the number of tiles in the lattice in one direction. 
The $\delta(z)$ is the delta of Kronecker.

\section{Equilibrium analysis and critical behaviour}

The numerical simulations done in order to study the equilibrium 
characteristics have been performed using the parallel
 tempering algorithm \cite{giappo}.
We have chosen open boundary conditions on the two dimensional lattice.
Indeed using periodic boundary conditions the fact that periodic tilings
do not exist would have implied that the energy of the ground state
would have been different from zero.
With open boundary conditions  the energy of the ground state is,
by definition of (\ref {hamil}), equal to zero.
It is equivalent to say that all the tiles are edge by edge perfectly 
matched,  forming some non-periodic structure.  

\subsection{Phase transition}

At different temperatures, for every system we have  computed 
the energy and the specific heath. 
Numerical simulations have been brought about
 on square lattices of different size:
from a linear size of 8 to one of 32.
Every equilibrium simulation has been carried out for a number of MC steps 
going from 10 millions to 100 millions, relatively to the size of the lattice.
A range of temperature between 0 and 5 has been observed, then concentrating 
in small intervals in what comes out to be the critical region, around
$T=0.4$.
We controlled that at zero temperature the energy goes to zero and that
the specific  heat computed from the derivative
 of the energy coincides with the one computed from the energy
fluctuations. These are checks of a correct thermalization.
\begin{figure}
\begin{center}
\label{fig:calori}
\includegraphics[width=0.49\textwidth , height=0.40\textwidth]{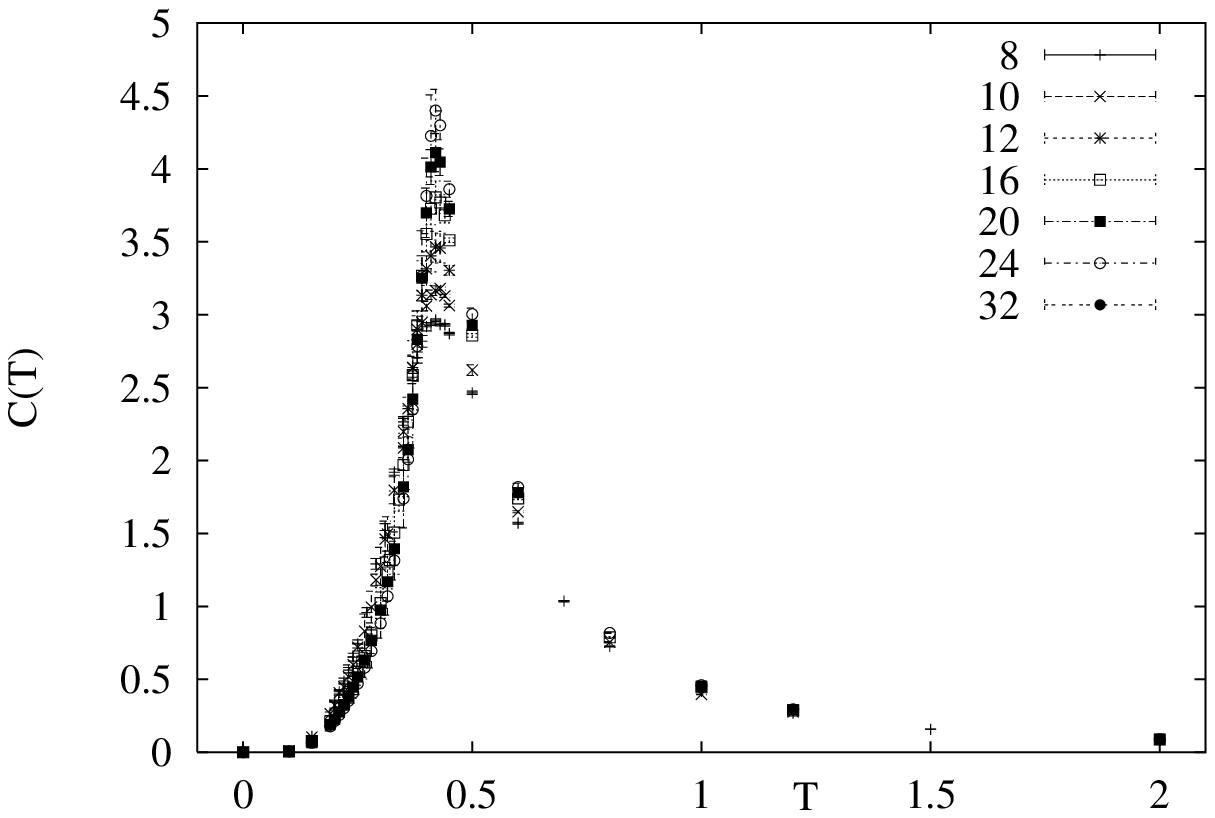}
\includegraphics[width=0.49\textwidth , height=0.40\textwidth]{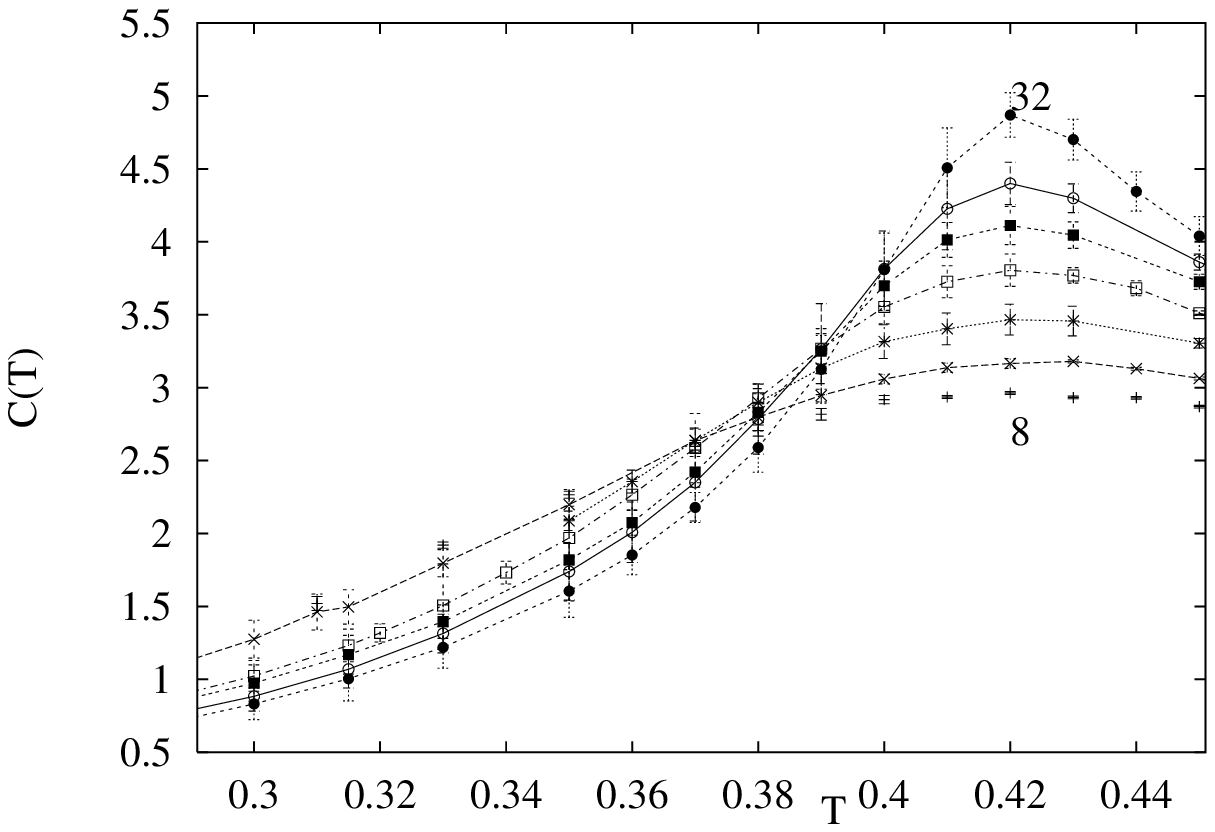}
\caption{\small {
Left: specific heats in the whole  probed region. Right: specific heats 
 around the critical region. Linear sizes from $8$ 
to $32$ are plotted. The crossing point shift with the size of the system
following a FSS.
We can observe both the raising of the slope of the
curves around the crossing point and  a growing peak at $T=0.42$}}
\end{center}
\end{figure}

The specific heat presents a smeared but clear
change around the temperature $0.4$ (see figure \ref{fig:calori}), from 
a lower value for lesser temperatures to a higher value for greater.
Increasing the size of the system we can observe that
the crossing points between two curves of different size moves towards right 
and that the slope of the
$C_L(T)$ line increases in this interval of temperatures as $L$ increases, 
but we also see that a peak grows around 
$T=0.42$. 

Performing Finite Size Scaling (FSS) analysis of the crossing
points of the specific heat curves we find that their abscissa 
tends to a $T_{cross}(\infty)= 0.398 \pm 0.007$ as $L\fd \infty$, with 
an exponent $\nu=1.6 \pm 0.5$
 and that the  behaviour of 
the slope with increasing size is compatible with a diverging fit at that
temperature. 
In this case we would have a jump at $T_c=T_{cross}(\infty)$,
which corresponds to a critical exponent $\alpha$ equal to zero.

But at the same time  the fit  of the  peak height is consistent with
 a divergence at 
$T_c=0.42 \pm 0.01$. Knowing that at criticality 
$C \sim |T-T_c|^{-\alpha}$ and 
$\xi \sim |T-T_c|^{-\nu}\sim L$ we  get $C\sim L^{\alpha/\nu}$.
Our data are both consistent with  a power-law divergence  ($\alpha/\nu = 0.35 \pm 0.01$),
and with a logarithmic one ($\alpha=0$).
A similar value of the critical temperature 
was found, in this last hypothesis, by Janowsky and  Koch  
\cite{koch}.

In any case there is evidence for a second order phase transition.

\subsection{Order parameter}

Wang's tiles satisfy a particular property: if all the tiles of the two
 contiguous {\it west} and {\it south} sides (or equally of the 
{\it north} and {\it east} sides)
 of the  square lattice are fixed, then at most one aperiodic 
tiling can be formed. This is due to the fact that each tile of the Wang set
has different combinations of west-south (or north-east) colours. Then at most 
only one  of them can be put,
 in the bottom-left (either top-right) corner.
 The same is true for the tiles to be
 put in the two corners created by putting the first tile. If any, there will
 be only one combination available also for them.
Of course this is valid at $T=0$ in our system.

This condition implies that ground state degeneracy can not increase
 faster than $16^{2L-1}$and therefore the entropy density is zero 
at zero temperature:
$s_L(0)\sim a/L$, where $a < 8\log 2$. Since just a
few north-south or west-east combination are allowed the actual constant
in the entropy is smaller than $8 \log 2$ (a stricter upper bound
of $a=\log 12=3.585\log 2$  can be easily obtained).

In order to identify an order parameter we have to break this degeneracy.

We have then simulated the parallel evolution of two copies of the system,
with the following procedure: one system reaches the equilibrium,
then a copy of the equilibrium configuration is done and the evolution of 
this second copy is performed. One foundamental
constraint is set: the boundary tiles on the south and east  
 sides of the lattice stay unchanged in the dynamics 
towards equilibrium of the second copy.

 For $T>0$ the equilibrium 
states are no more the exactly matched tilings. Because of the thermal noise
some couple of neighbours sides of the Wang's tiles can be unmatched. 
Furthermore  there won't be
a unique tiling minimizing the energy. The
  second copy  can then evolve to a different configuration.

We are interested in looking what happens when the temperature is increased 
from below to above the critical temperature,
and how the behaviour  is sensitive to the size of the system.
With this aim we introduce an {\it{overlap}} that depends on the distance $l$
of the tiles from the two contiguous boundary sides that are fixed once that
the first copy has reached equilibrium:
\be
q(l)= \frac{1}{2(L-l)+1}
\sum_{y\geq l, x=l} \sum_{x\geq l, y=l}
{\overline {\left<\delta\left(T^{(1)}_{x,y}-T^{(2)}_{x,y}\right) \right>}}
\ee
Where the average ${\overline{ (...) }}$ has been performed 
over different realizations 
of the configurations, $\left<(...)\right>$ is the time average at equilibrium
 and $l=1,...,L$.

The overlap $q(l)$ is computed  along the diagonal:  starting 
from the vertex shared by  the two fixed contiguous sides, $q(1)$,
and ending at  the opposite vertex of the lattice, $q(L)$.
In order to gain more statistics the overlap is built averaging over 
the $2(L-l)+1$ elements in the row 
of ordinate $l$ with the coordinate $x\geq l$
and  in the column of abscissa $l$ with $y \geq l$
and assigning  this average value to the  'diagonal' function $q(l)$.

\begin{figure}[!htb]
\begin{center}
 \begin{tabular}{rl}
   \includegraphics[width=0.49\textwidth,height = 0.3\textheight]{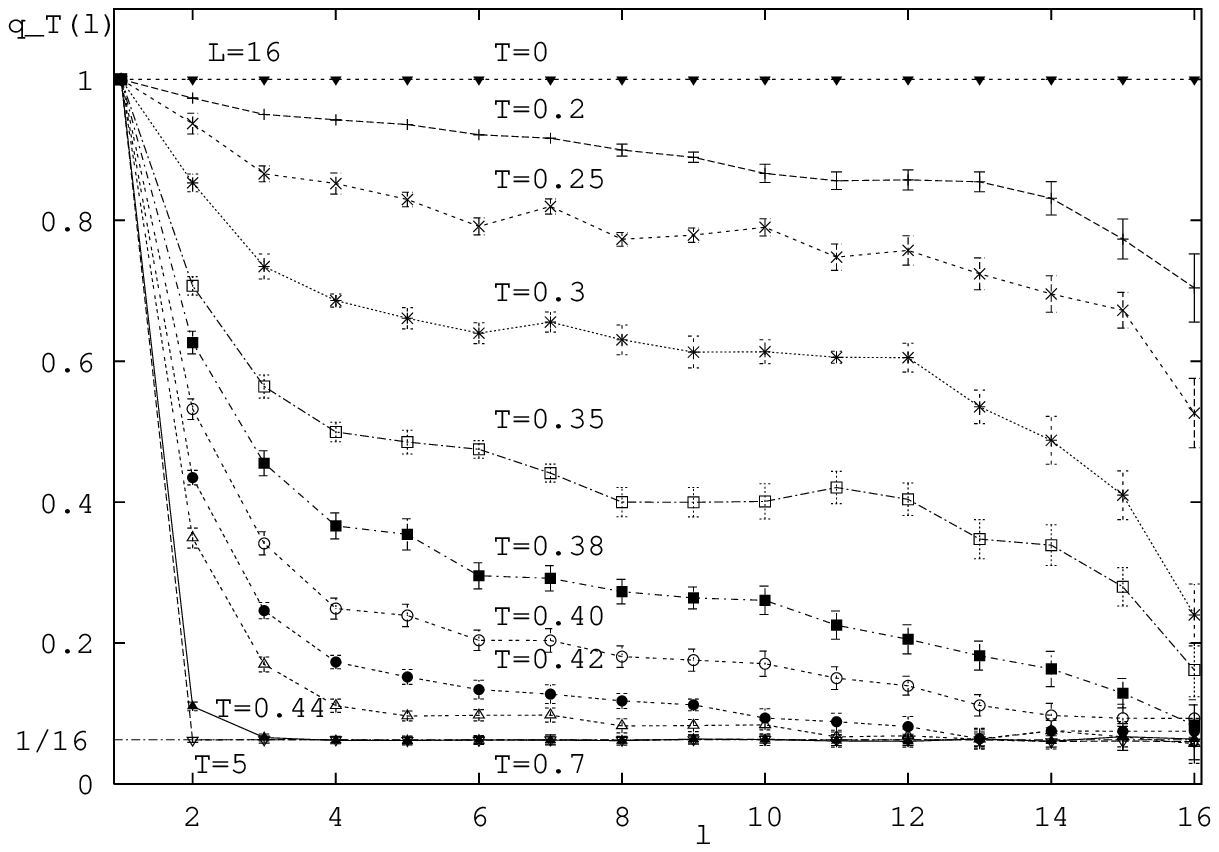}
  & \includegraphics[width=0.49\textwidth,height = 0.3\textheight]{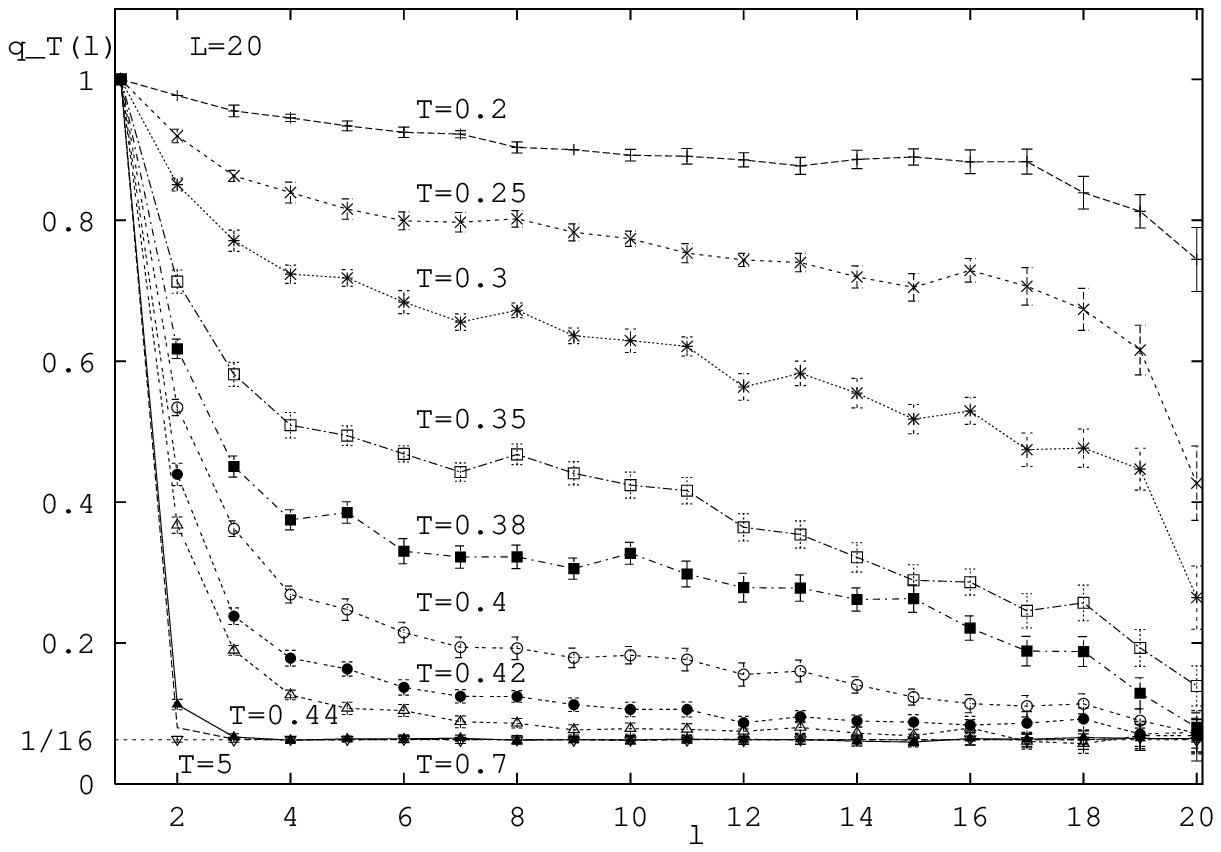}\\
\vspace{-0.5 cm}
  \includegraphics[width=0.49\textwidth,height = 0.3\textheight]{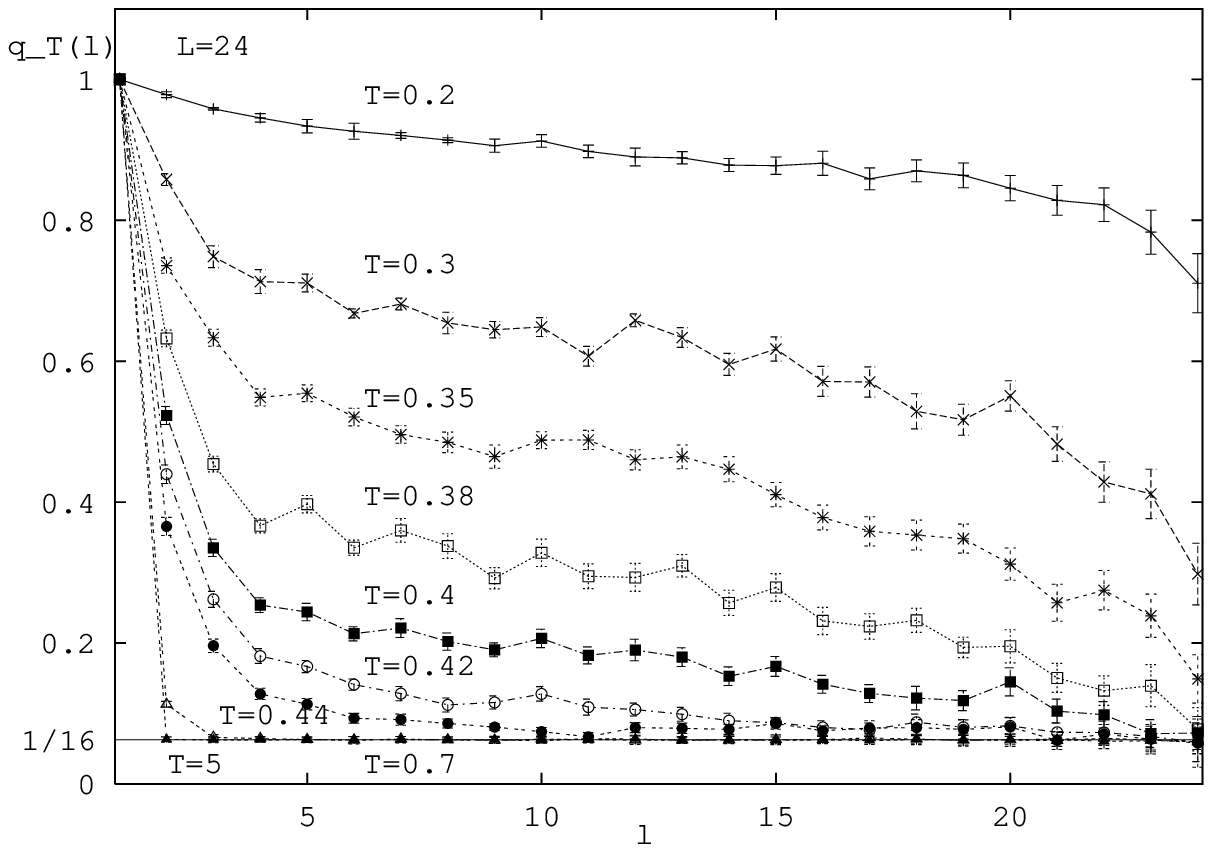}
  &
  \includegraphics[width=0.49\textwidth,height = 0.3\textheight]{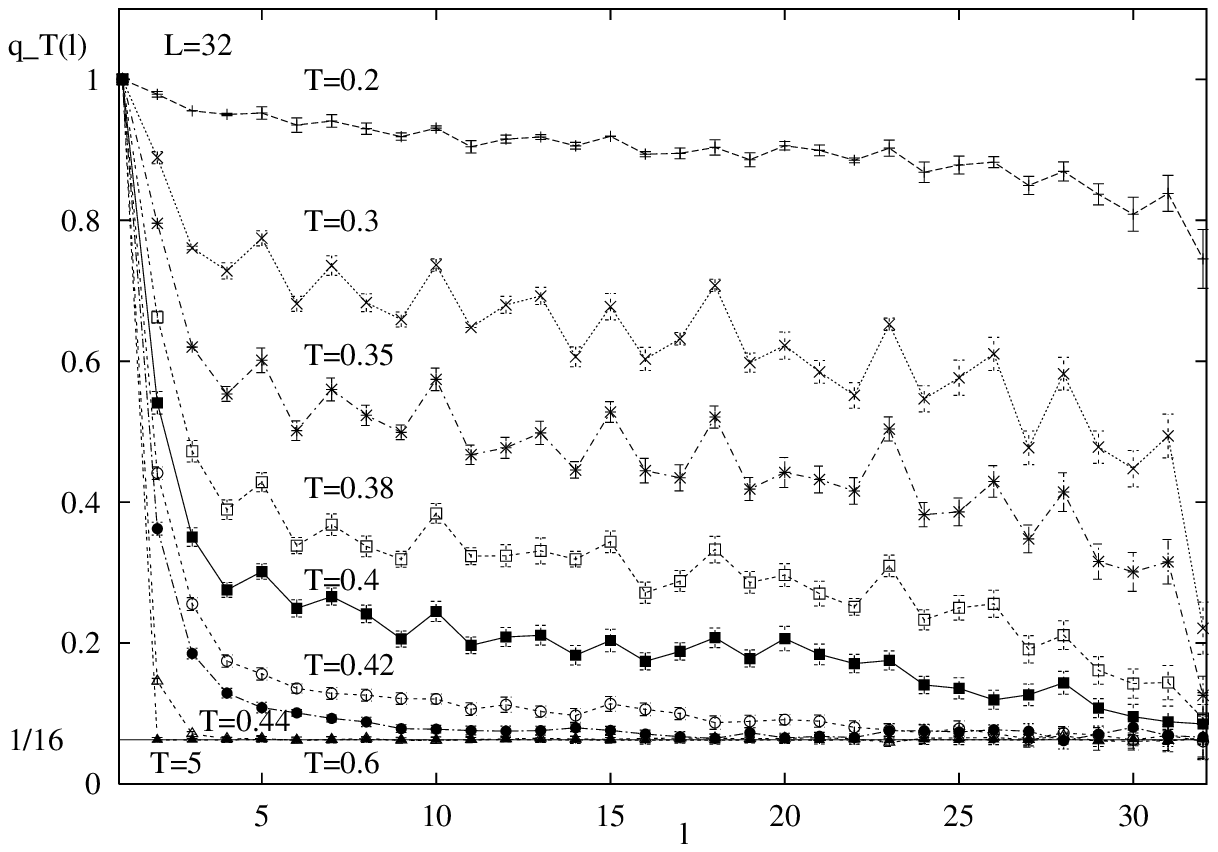}
   \end{tabular}
   \protect\caption{{\small{$q(l)$ profiles at different temperatures and 
different sizes. $T$ goes 
from 0.2  above to 5.0 at the bottom. 
The behaviours for $L=16, 20, 24, 32$ are shown.
For $L=16$ also the behaviour at $T=10^{-9}$ is plotted. }}}
   \label{fig:qlT}
\end{center}
\end{figure}

The statistic sample is formed repeating the simulation  several times
with different initial configurations, in order to obtain different 
equilibrium configurations (we have always checked if the same 
equilibrium configuration could possibly
appear more than once starting from different initial configurations
to avoid annoying biases of the statistic sample, but this never happened). 
Every size has been simulated for at least 100 different initial
configurations.

 At  zero temperature the overlap is always  one.

 At higher temperature but  below the phase transition
it   goes to a constant lesser than one 
(far enough from the boundary $L$, where finite size effects are overwhelming).

Above the transition point $q(l)$  decays very rapidly, compatible with 
an exponential, to the lowest  possible value, corresponding to completely 
uncorrelated copies ({\it hot} or {\it disordered} phase).
Since the tiles are of sixteen different types and the probability distribution
of the tile-types is uniform, this lowest value is equal to $1/16$.
  
In figure \ref{fig:qlT} we show the behaviour of $q(l)$ at different
temperatures, both above and below the critical one, for different sizes.
From this figures we can see that the $q(l)$ seems to 
approach some {\it plateau} for
$l\sim L/2$ at temperatures below $T\sim 0.42$. 
This is more evident as we go to bigger sizes.

\begin{figure}[!htb]
\begin{center}
  \includegraphics[width=0.8\textwidth,height = 0.3\textheight]{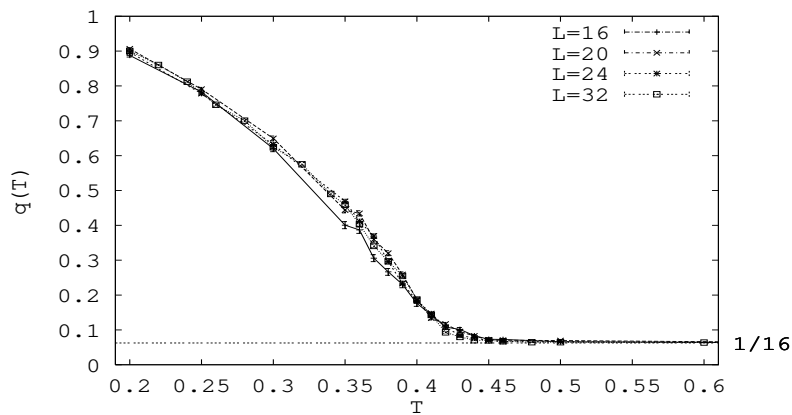}

   \protect\caption{{\small{$q_L(T)$ profiles for $L=16, 20, 24$ and $32$. 
$q_L(T)$ 
is the plateau value computed respectively on intervals of $3$, $4$, $6$ 
and $10$ tiles  }}}
   \label{fig:qLT}
\end{center}
\end{figure}

We can compute a $q(T)$ for every size, averaging over the 
{\it plateau} values of $q(l)$. The plateau is every time chosen taking a small
window around $L/2$ and then enlarging it as long as the {\it plateau}
 value stays constant.
As $l$ grows further finite size effects destroy 
this plateau. We observe that this parameter shows a change approaching 
$T_c$: when $T<T_c$, it increases
 from the value of $1/16$ that it has at high temperature to the value
of 1 that it reaches at zero temperature.
The profiles of $q_L(T)$ are plotted in figures \ref{fig:qLT}.

\begin{figure}[!htb]
\begin{center}

  \includegraphics[width=0.8\textwidth,height = 0.3\textheight]{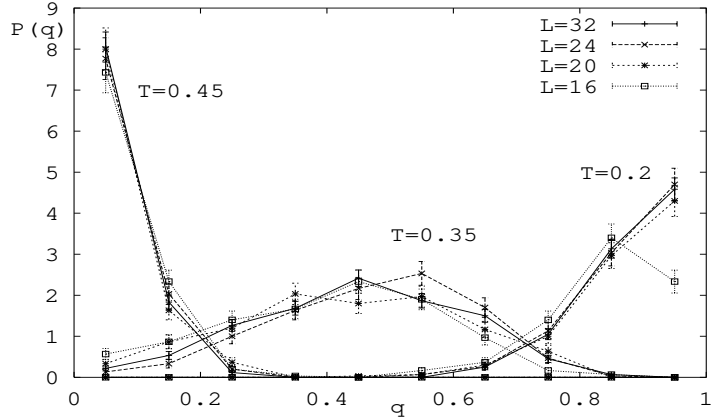}
   \protect\caption{{\small{Probability distribution of the overlap's values
for three different temperature: $T=0.45$ is already in the disordered
phase, while $T=0.35$ and $T=0.2$ are under the critical temperature. 
The behaviour is slightly dependent on temperature. For $T\fd 0$ the distribution tends to a $\delta$ function on the value 1.}}}
   \label{fig:pq}
\end{center}
\end{figure}

\begin{figure}[!htb]
\begin{center}

  \includegraphics[width=0.17\textwidth,height = 0.1\textheight]{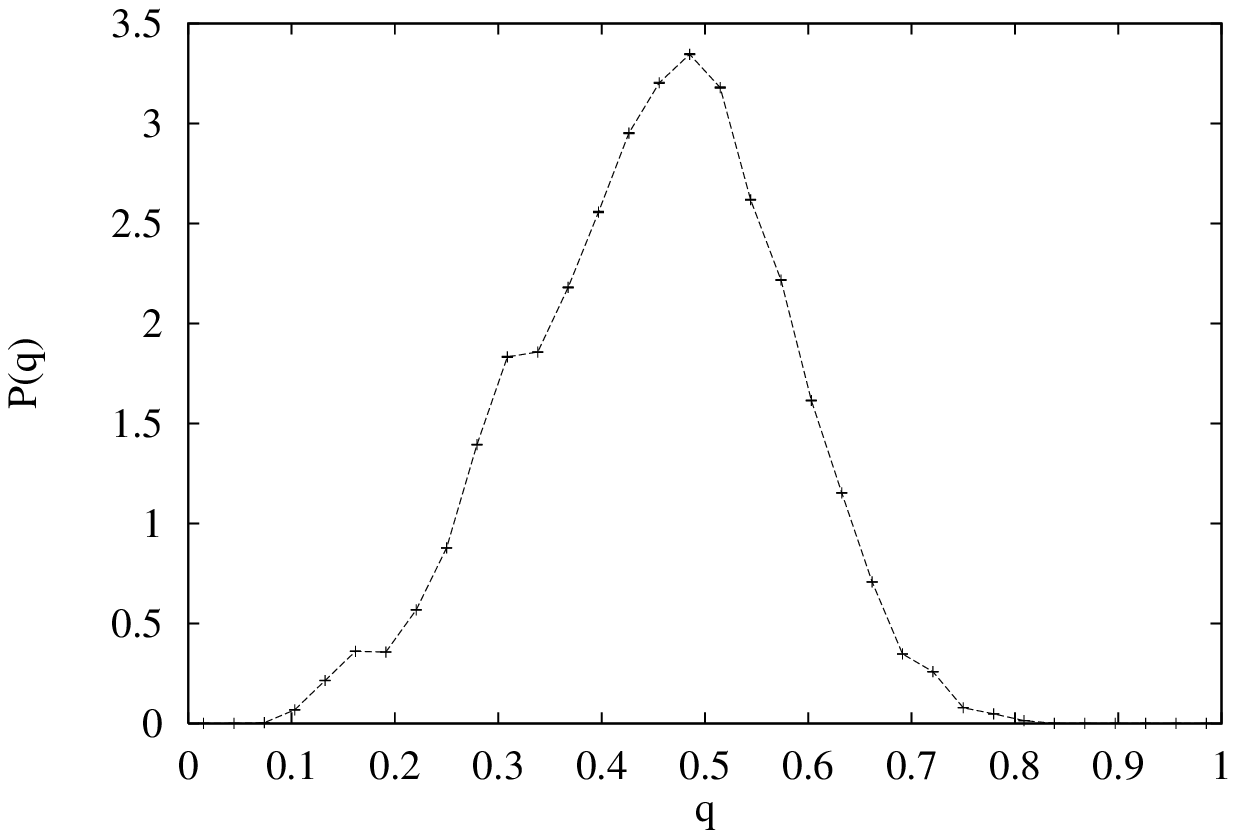}
  \includegraphics[width=0.17\textwidth,height = 0.1\textheight]{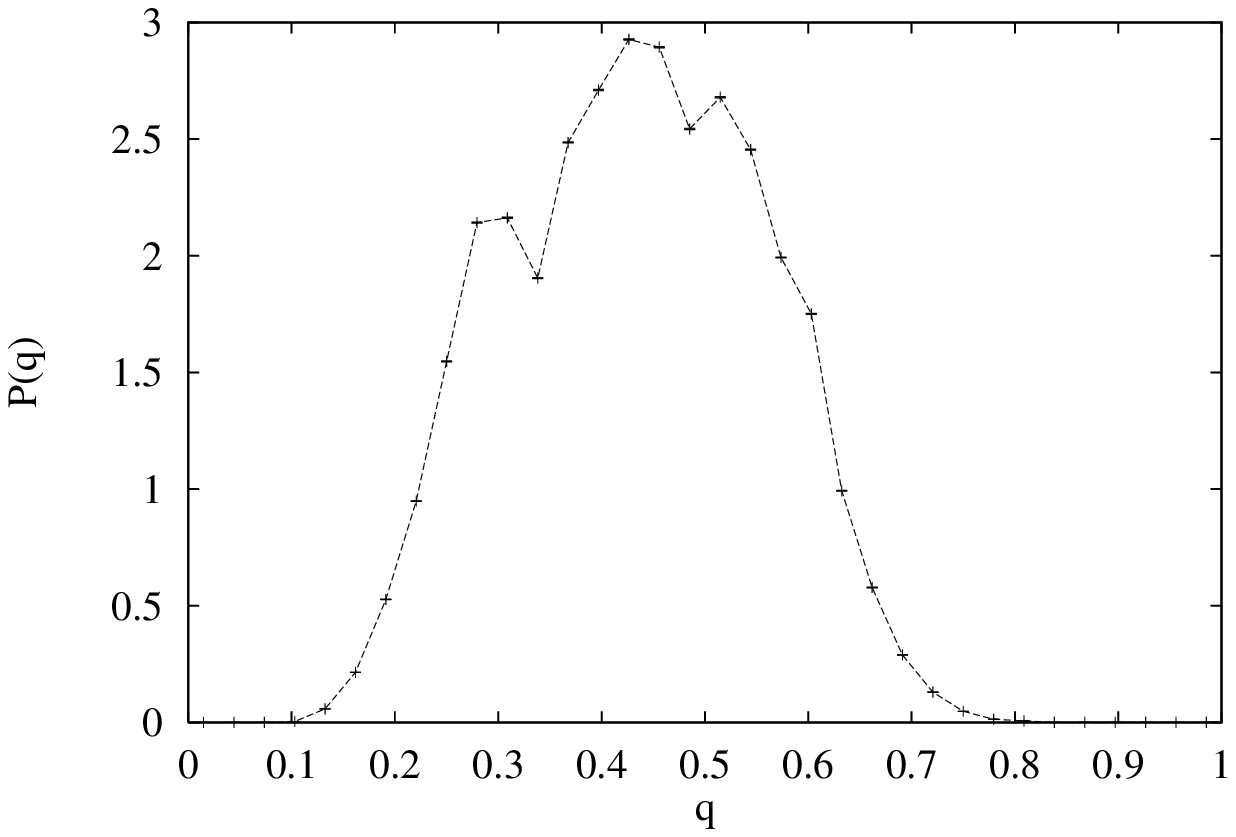}
  \includegraphics[width=0.17\textwidth,height = 0.1\textheight]{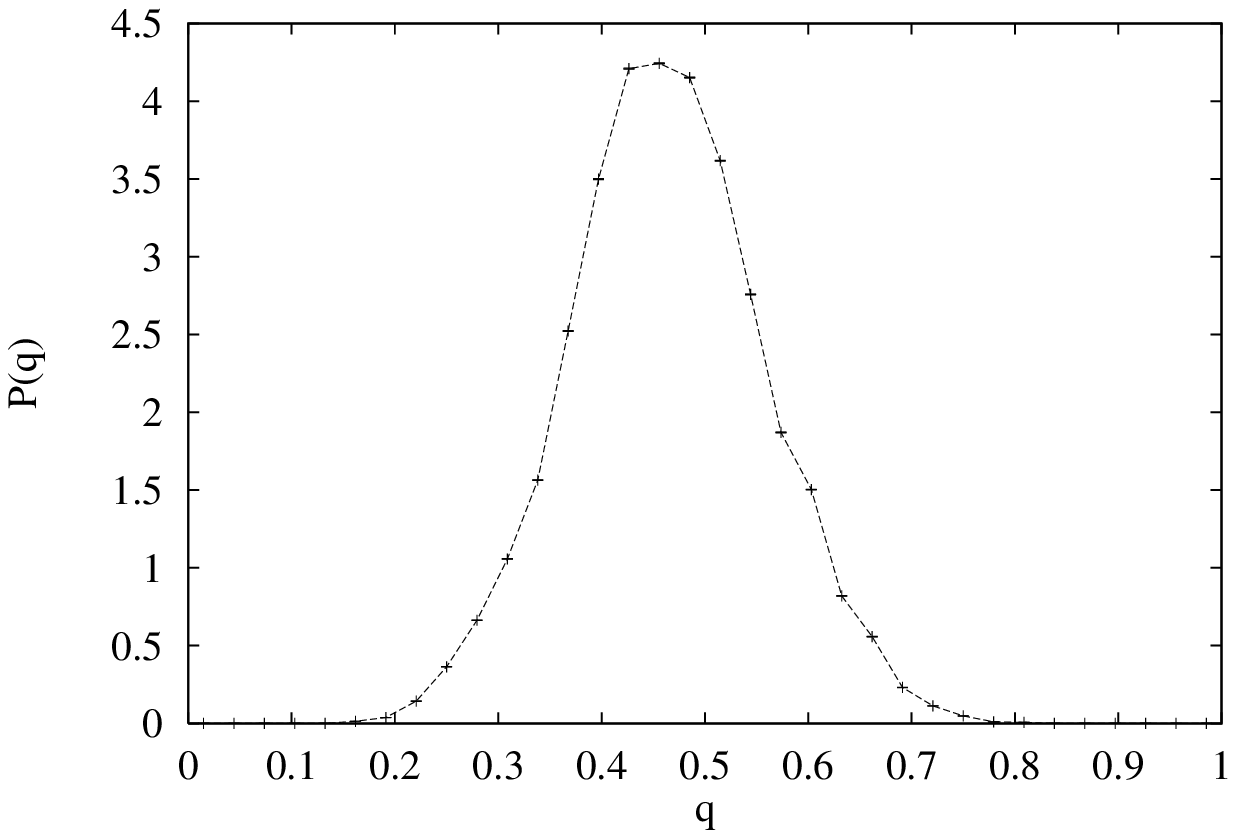}
  \includegraphics[width=0.17\textwidth,height = 0.1\textheight]{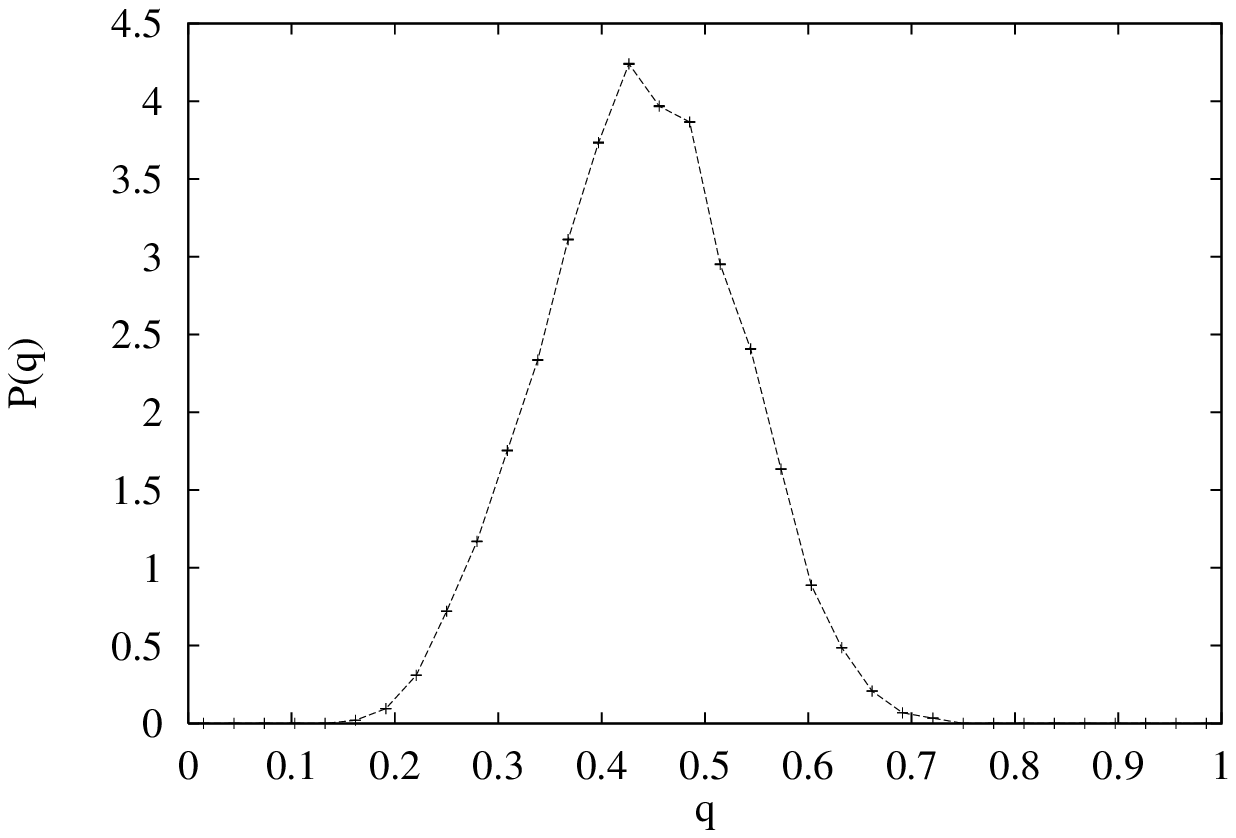}
  \includegraphics[width=0.17\textwidth,height = 0.1\textheight]{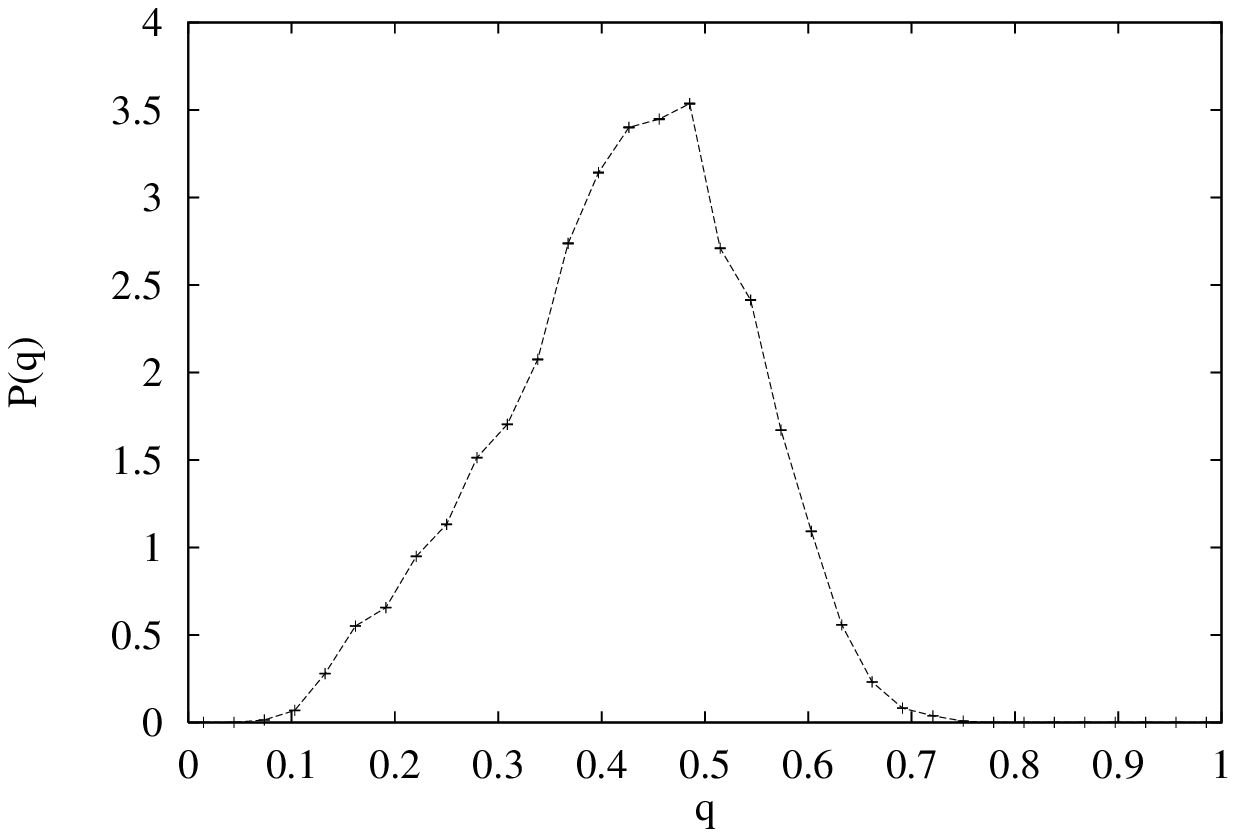}

   \protect\caption{{\small{Different profiles for different realizations
	of the initial conditions at $T=0.35$ 
	for a system of linear size 32.}}}
   \label{fig:profili}
\end{center}
\end{figure}

The probability distribution of the overlap's values at the plateau has a 
non-trivial shape as soon as $T<T_c$ (figure \ref{fig:pq}). 
This is not so astonishing since fixing 
the boundary conditions we have broken the degeneracy of the equilibrium 
states. (We stress that due 
do the overlap's definition the distribution doesn't show any
 symmetry $q\fd -q$).
The form of the $P(q)$ averaged over different realizations of the tile's
configurations at two continuous edges of the lattice 
is not strongly dependent from the size of the samples, at least for the 
simulated cases.

The probability distributions of the overlap for different realizations
of the  equilibrium boundary conditions in the bottom and left edges of the
tiling are represented in figure \ref{fig:profili}.

\section{Off-equilibrium analysis}

In order to study systems with very slow relaxation times, much greater than
the experimental times, it is very important to look at the
 off-equilibrium dynamics. This approach describes  glasses, 
 spin-glasses and, in general, any system with many equilibrium states
divided by high free energy barriers,
in a more realistic way in comparison with an equilibrium approach,
since equilibrium is, in practice, never reached during experiments.

To keep the sample out of equilibrium during the run of a numerical simulation
we need the typical distance $\xi(t)$ 
over which the system has reached equilibrium
 at a certain time to be  always smaller than the linear size $L$ 
of the lattice.
This distance is also called {\it dynamical correlation distance}.
Practically, in order to be sure 
of avoiding thermalization, we choose sizes bigger than those 
used in the static analysis. 
The greater is the size the longer is the time needed to the 
correlation distance  to 
reach the size of the system.
More over we used in this case a standard Monte Carlo algorithm instead 
of the parallel tempering, 
so that the thermalization times increased sensitively
 under the transition point.

In order to study the response  of the system to an external field
we can  embed the system  into a perturbative field.
 Namely a field directed in one of the sixteen
possible directions in the 'tile-types space', where a particular tile 
of the set of sixteen can be
favoured (positive field) or disfavoured (negative field).
This  can be a uniform or non-uniform field. 
To avoid any preference towards one particular type of tile we have chosen
a    non-uniform random field, whose 
value at each  site is independent from the other sites.

Thus we add to the Hamiltonian (\ref{hamil}) the perturbative term:
\be
\sum_{(x,y)}h^{sign}_{x,y}\left(1-\delta\left(T_{x,y}-h^{type}_{x,y}
\right)\right).
\ee
\noindent where
 $h_{x,y}=h^{type}_{x,y}h^{sign}_{x,y} $
 is the random external field pointing along one of the 
tile-types or opposite to it. 

\begin{figure}
\begin{center}
\includegraphics[width=0.65\textwidth , height=0.3\textwidth]{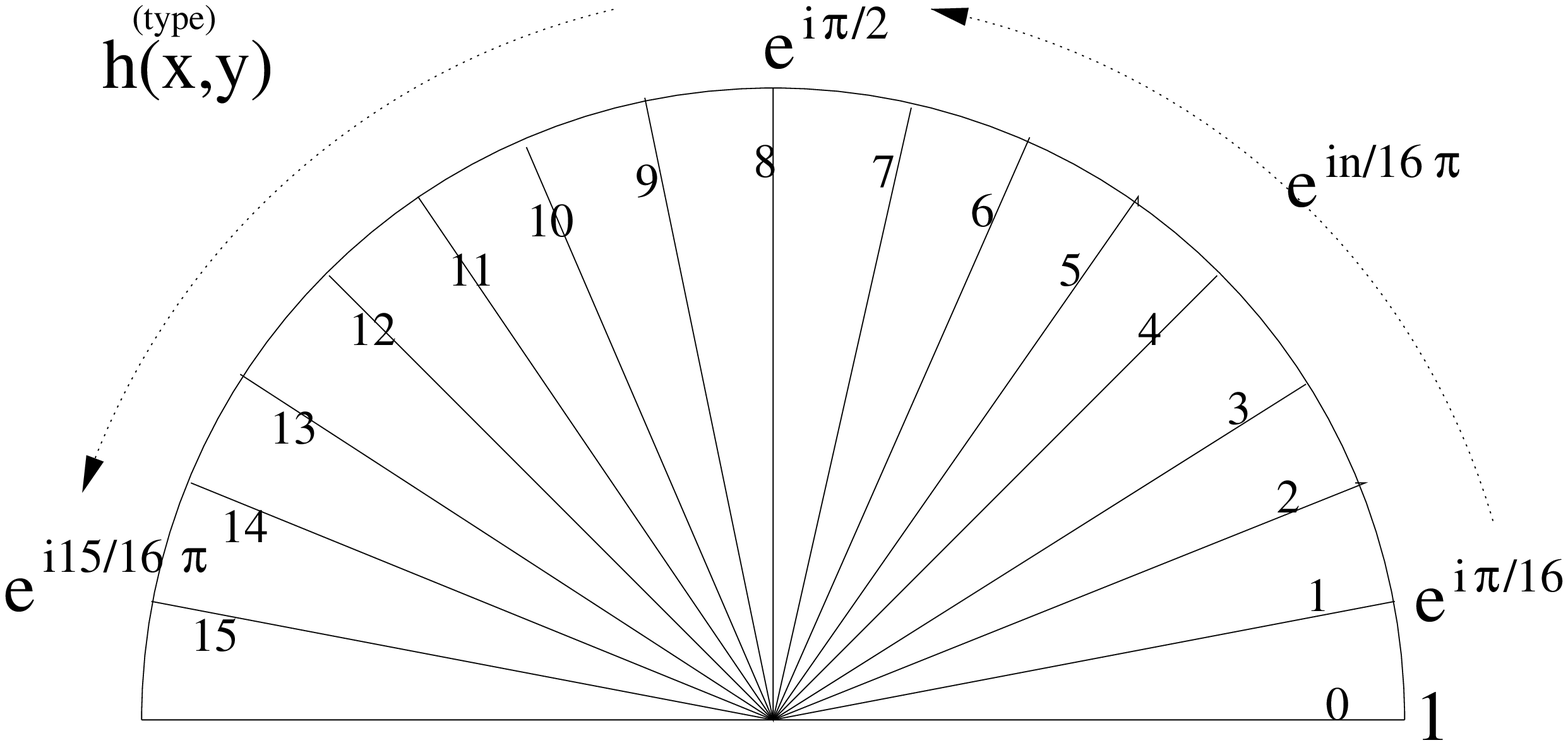}
\caption{\small{'directions' of the random field $h^{type}$ in the space 
of tile types, represented in the complex plane.}}
\label{fig:accadir}
\end{center}
\end{figure}

$h^{type}_{x,y}$ has
a uniform distribution of  the sixteen possible choices of tile-type,
and $h^{sign}_{x,y}$ gives the magnitude of the field and its sign
(it can be
randomly positive or negative, always according to a uniform distribution).%

The probability distribution can then be written as:
\be
{\cal{P}}(h_{x,y})=\frac{1}{16}\sum_{n}^{0,15}
\delta \left(h_{x,y}^{type}-e^{i\pi\frac{n}{16}} \right) \times
\frac{1}{2}
\left(
\delta(h_{x,y}^{sign}-h_o)+
\delta(h_{x,y}^{sign}+h_o)
\right)
\ee
with $\overline{h_{x,y}}=0$ and $\overline{h^2_{x,y}}=h_o^2$.
$n=0,...,15$ gives the  'direction' of the field in a 
representation in the complex plane (see figure \ref{fig:accadir})

Once the system is cooled from high temperature, it is left evolving 
until a certain time $t_w$, usually called {\it waiting time}.
At $t=t_w$ the field is switched on and we begin recording the values of the 
temporal correlation function $C(t,t_w)$ and of the integrated response 
function $m(t,t_w)$.

For our model they are defined as follows:

\be
C(t,t_w)=\frac{1}{L^2}\sum_{x,y} \delta(T_{x,y}(t)-T_{x,y}(t_w)),
\ee
\be
m(t,t_w;h)=
\frac{1}{L^2}\sum_{x,y}\frac{
{\overline{ \left< h^{sign}_{x,y}(t_w)\ \ \delta(T_{x,y}(t)-
h^{type}_{x,y}(t_w))\right>}}}{h_o}
\label{linear}
\ee
and the susceptibility is
\be
\chi(t,t_w)=\lim_{h_o \fd 0} \frac{m(t,t_w;h)}{h_o}.
\ee
For numerical computation it becomes:
\be 
\chi(t,t_w)\sim \frac{m(t,t_w;h)}{h_o}.
\ee
Here  $\left<(...)\right>$
 is the average over different dynamical  processes and
${\overline{(...)}}$ the average over the random realizations of the 
perturbative external field.
In a system at equilibrium the Fluctuation-Dissipation Theorem (FDT) holds:

\be
\chi(t-t_w)=\frac{1-C(t-t_w)}{T}.
\ee
\noindent where we have made explicit use of the fact that the correlation 
function is defined in such a way that $C(t_w,t_w)=C(0)=1$.

Out of equilibrium this theorem is no more valid.
Anyway a generalization is possible \cite{bouchaud}, 
at least in the early times of the dynamics. This generalization is made
introducing a multiplicative factor $X(t,t_w)$ depending on two times 
such that :
\be
\chi(t,t_w)=\frac{X(t,t_w)}{T}\left(1-C(t,t_w)\right)
\ee
In a certain regime, called {\it {aging}} regime, $ X(t,t_w)<1$ and 
the FDT is violated.
An important   assumption is that this modified coefficient $X(t,t_w)/T$
depends on $t$ and $t_w$ only through $C(t,t_w)$.
Its inverse is also called {\it effective temperature} 
$T_{e}\equiv T/X\left[C(t,t_w)\right]$ since the system in this regime,
for a given time-scale, seems
to behave  like a system in equilibrium at a  temperature different from the 
heat-bath temperature.

In terms of susceptibility and correlation functions we can write
a general functional dependence:
\be
\chi(t,t_w)=\frac{1}{T} S\left[C(t,t_w)\right]
\ee
 
The $S[C]$ here defined 
would  be  $1-C$ if we were at equilibrium.

From our probe we find that we first have a regime where the relation
is linear  with a coefficient equal to the heat-bath temperature.
This early regime is sometimes called {\it {stationary}}, since the 
observables computed at very short times (compared with $t_w$)
do not depend on the age of the system. During this regime the
system goes fast towards a local minimum. 
\begin{figure}[!htb]
\begin{center}
  \includegraphics[width=0.78\textwidth,height = 0.3\textheight]
{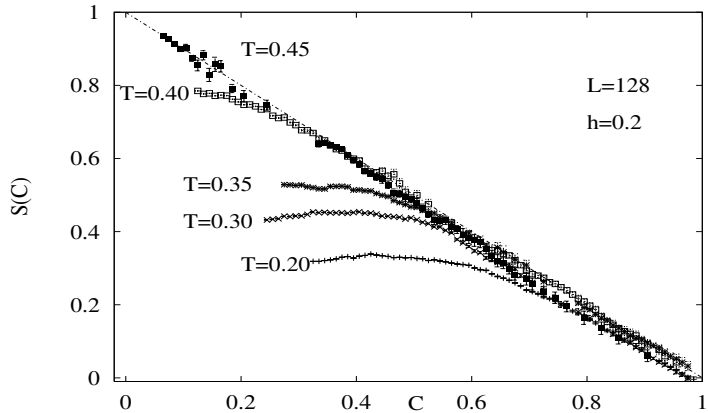}
 \protect\caption{\small{ L=128, h=0.2, S[C] at different temperatures.
$T=0.45$ is above the phase transition: in this case there is no regime
 in which the fluctuation-dissipation theorem does not hold. The time
 evolution of the system for $t\geq t_w$
 has to be read going from right ($C(t_w,t_w)=1$) to left.}}
  \label{fig:sc_varie1}
\end{center}
\end{figure}
\begin{figure}[!htb]
\begin{center}
 \hspace{-2 mm} 
    \includegraphics[width=0.49\textwidth,height = 0.3\textheight]
	{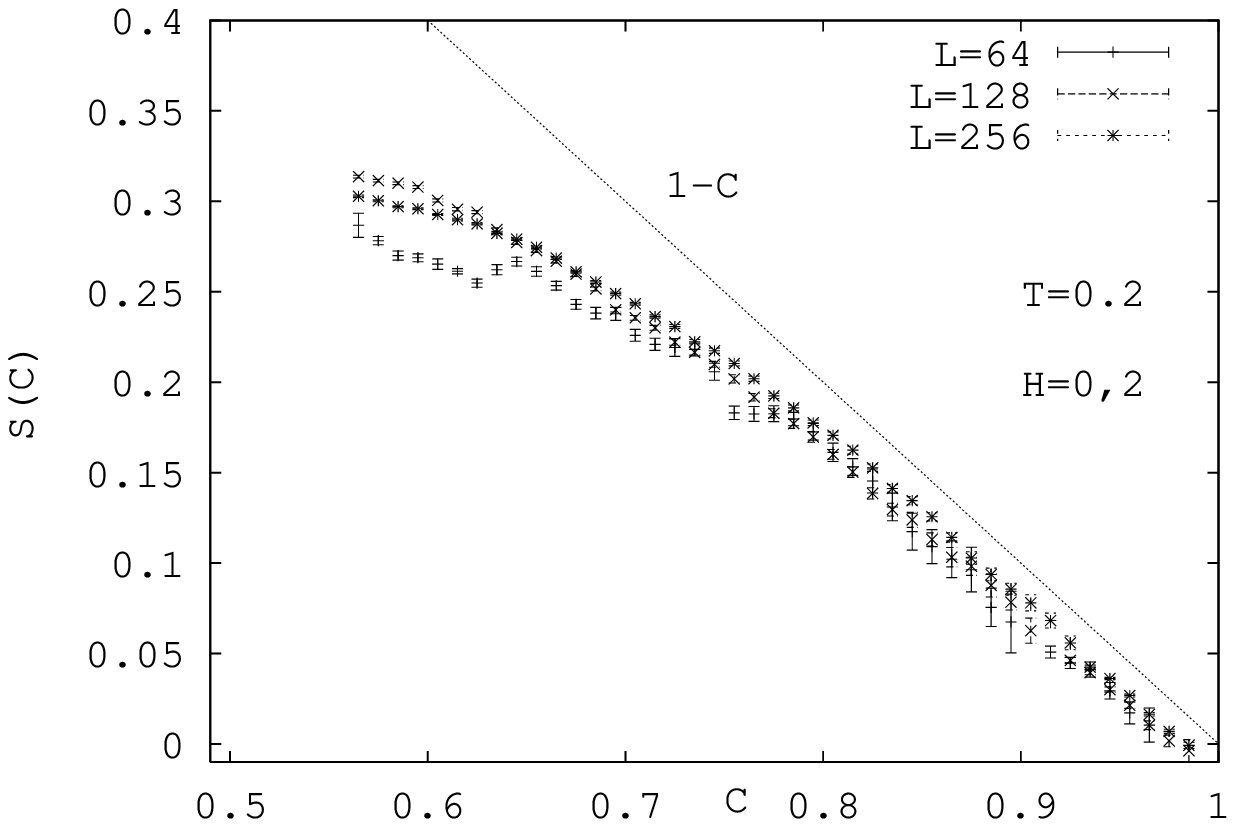}
    \includegraphics[width=0.49\textwidth,height = 0.3\textheight]
	{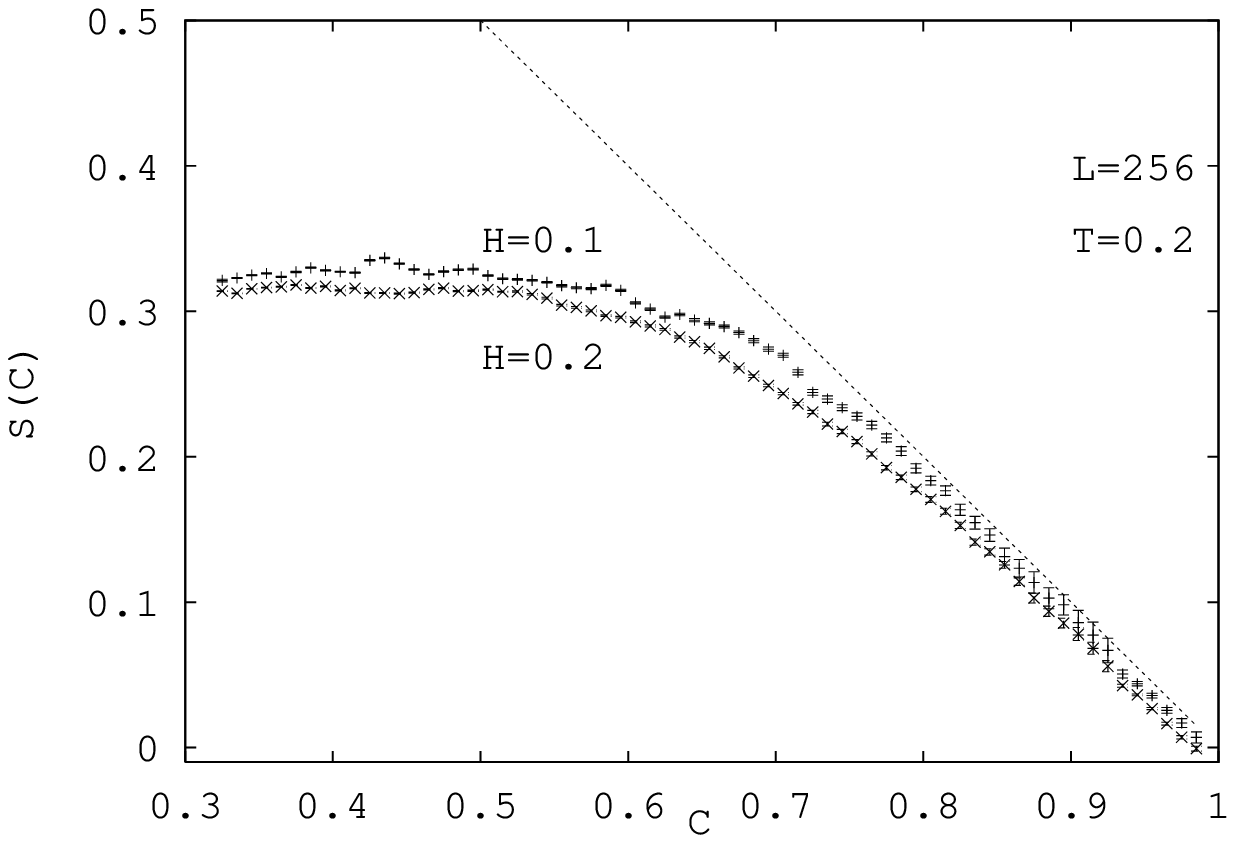}

    \protect\caption{\small{ Left: T=0.2, h=0.2, S[C] for different sizes.
	 The dependence from the size is little. Right: L=256, T=0.2, S[C] for 
	different values of $h$. }}
    \label{fig:sc_varie2}
\end{center}
\end{figure}

\begin{figure}[!htb]
\begin{center}
  \includegraphics[width=0.50\textwidth,height = 0.33\textheight]{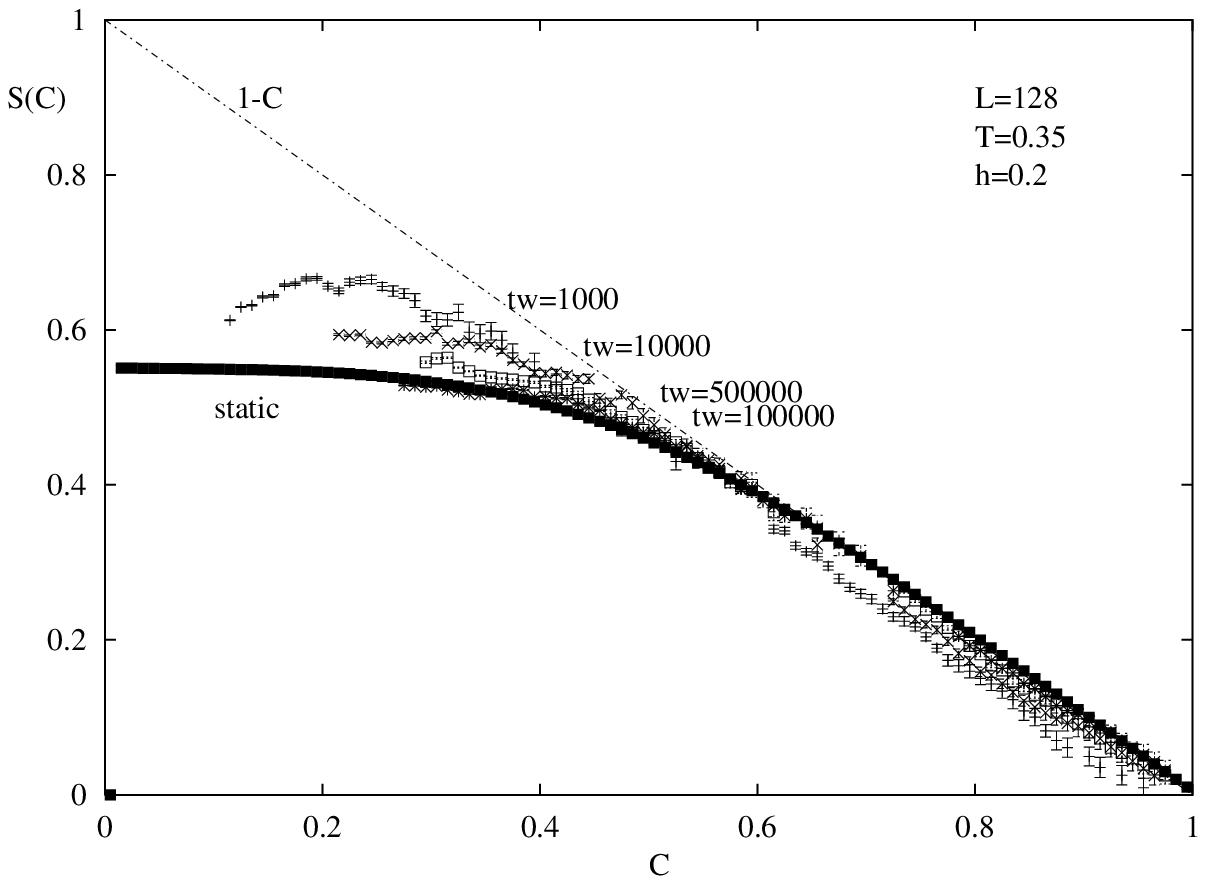}
  \includegraphics[width=0.48\textwidth,height = 0.33\textheight]{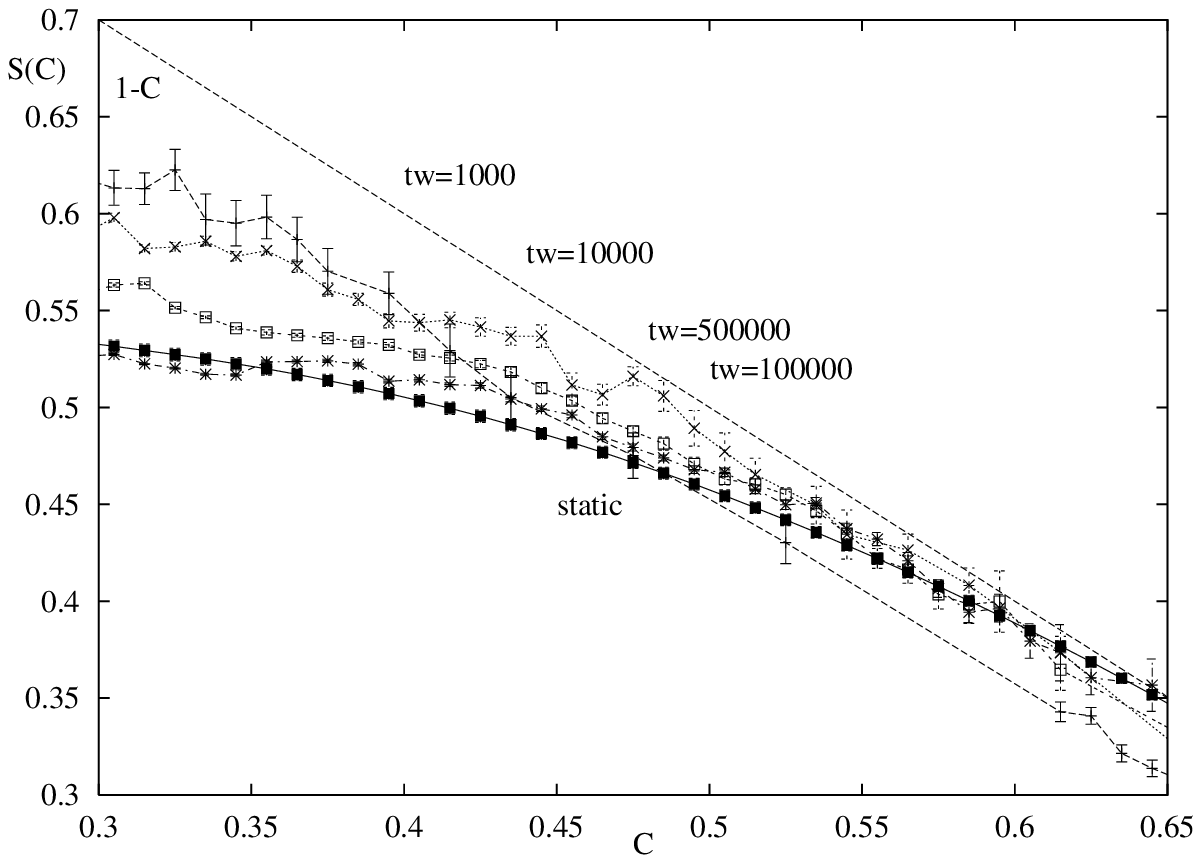}
  \includegraphics[width=0.50\textwidth,height = 0.33\textheight]{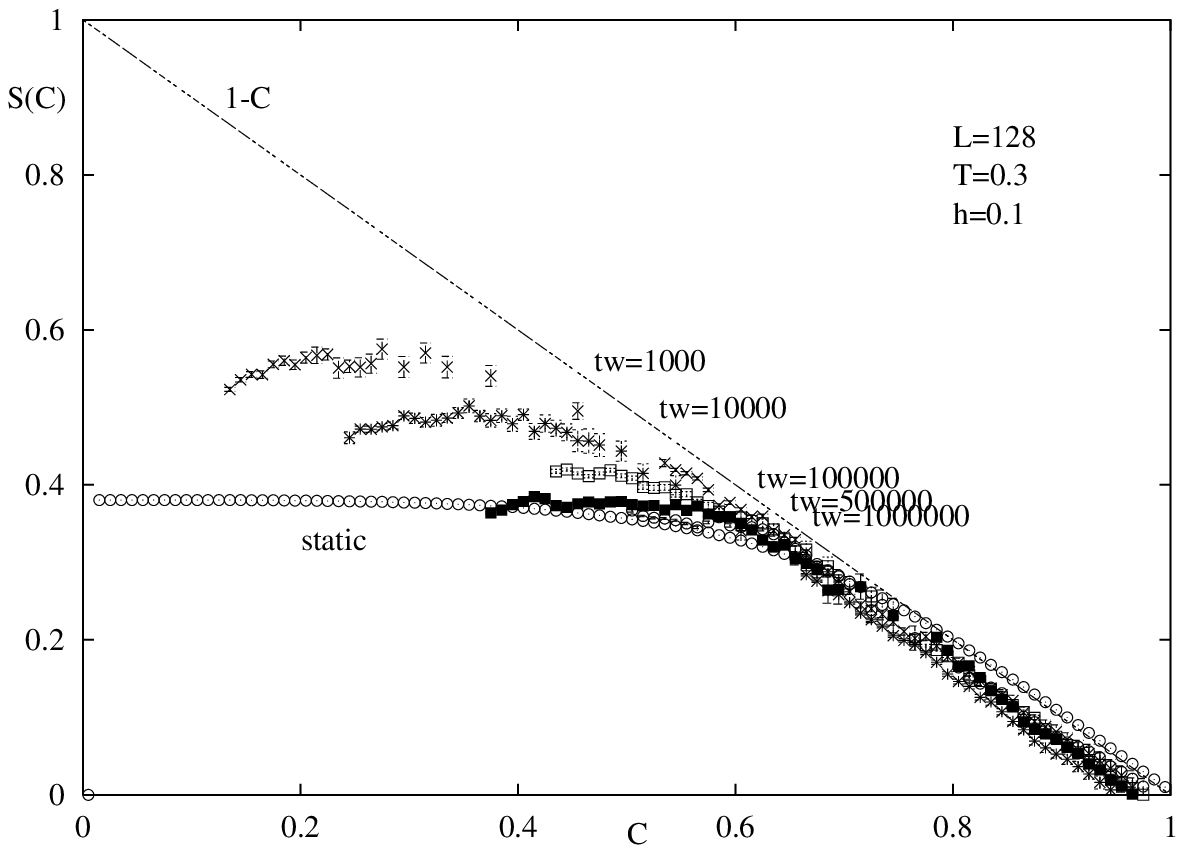}
  \includegraphics[width=0.48\textwidth,height = 0.33\textheight]{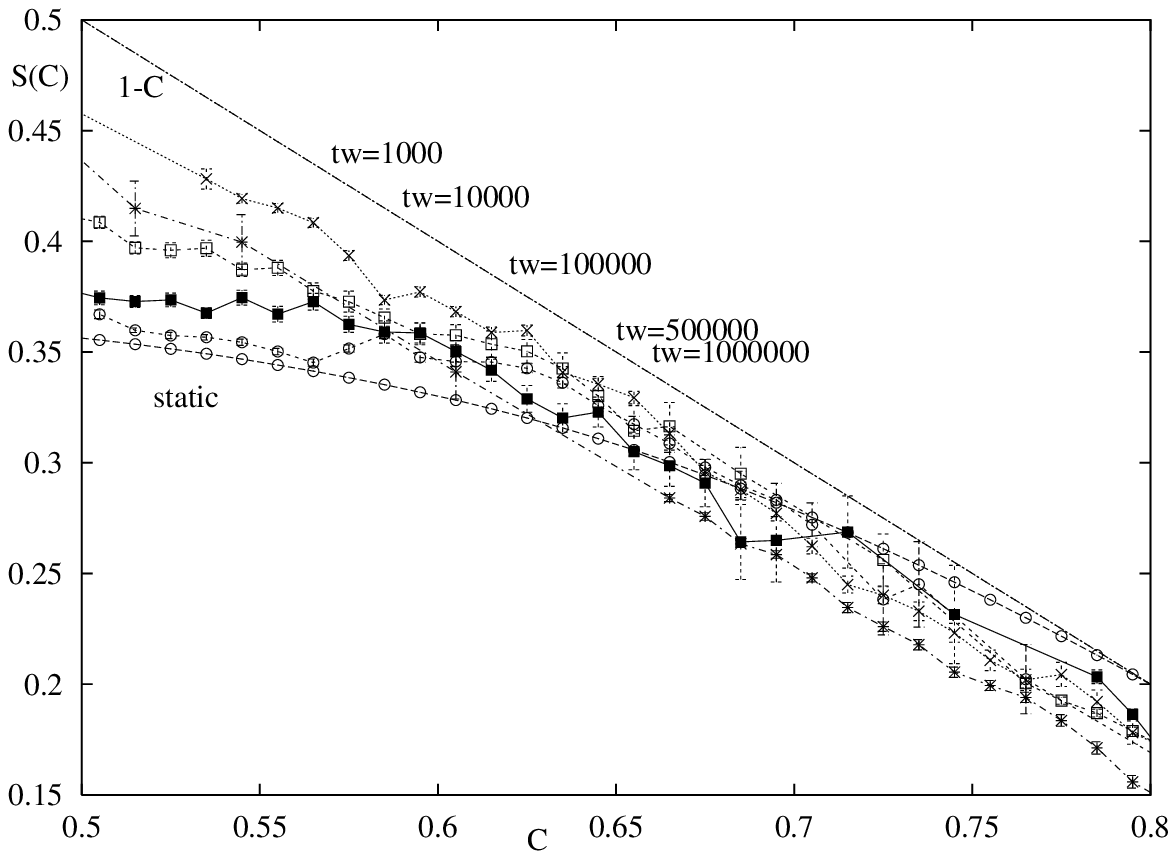}
   \protect\caption{\small{Comparison between $S[C]$ computed on 
dynamic  and  static data. The {\it static} curve is computed from the data
of a system of linear size $L=32$ at equilibrium. The other curves from the 
data of a system with $128 \times 128$ tiles out of equilibrium. Measurements 
starting at different waiting times $t_w$ are shown.
The two plots above represent the situation at $T=0.35$, where a 
perturbative external field of magnitude $h_o=0.2$ has been applied at $t_w$.
The two below show the relation between response and correlation functions 
at $T=0.30$, with $h_o=0.1$.
For large $t_w$ the dynamically determined $S[C]$ tend to lay on the asymptotic
{\it static} curve. On the rightside plots of
 the detail of the beginning of the bending is shown.}}
   \label{fig:sc_static}
\end{center}
\end{figure}

After this first time 
the $S[C]$ bends and the coefficient of $X(t,t_w)/T$ is no more 
the inverse heat-bath temperature. 
 $X(t,t_w)$ is now less
than one, as if the system would be at an effective temperature 
bigger than  the one of the heat-bath. 
It also seems to change continuously as the
system evolves, until it reaches zero (figures \ref{fig:sc_varie1} -
 \ref{fig:sc_static}).

The value of the autocorrelation
 function at which $S[C]$ leaves the $1-C$ line, equal 
to the average overlap value $q$ in the static,
 increases continuously with temperature, as we already observed in
the static analysis.

The $S[C]$ shows  no strong dependence from the magnitude of the 
field, 
at least for the values that we have used to perturbate the system.
Furthermore the system shows only a slight dependence from  the size 
 (measures on $L=64, 128, 256$) (fig. \ref{fig:sc_varie2}).

From the figures \ref{fig:sc_static} we can see that $S[C]$ moves towards some
asymptotic line  increasing  $t_w$.  

The initial difference between $1-C$ and $S[C]$ is due to lack of statistics:
if $N_{fr}$ is the number of field realization performed
in the simulation there is a difference of order $N_{fr}/L^2$ between 
the value of the integrated response function in the
 thermodynamic limit and the value
at finite size.

We can also look at the link with the static analysis.
Like in the case of mean field spin glasses we can suppose, following 
{\cite{cuku}}{\cite{mapa}}, that for 
$t,t_w \fd \infty$, $C(t,t_w) \fd q$ and  $X[C(t,t_w)] \fd x(q)$, where
$x(q)$ is the cumulative distribution of $P(q)$:
\be
x(q)=\int_{0}^{q} dq'\ P(q')
\ee
If this link is valid we can connect the susceptibility multiplied by the
heat-bath temperature at a certain correlation value \ \ ($S[C]$) \ \ to 
the integral \ \ $\int_{1}^{C} dq \ x(q)$.

In our case
there is an agreement between dynamic data and the values of this last 
integral computed on the static data. The two approaches are consistent
if the  external field is not too large, in order to avoid the 
non-linear effects
that are neglected in the derivation of formula {\ref{linear}}.
The field cannot  even be too small, though, in order 
to make the system move from
the meta-stable state in which it has gone during $t_w$.
The lower is the temperature at which we cool the system,
the bigger is the external field necessary to make it explore other parts 
of the space of states.
 A probe at too low temperatures then
 is not really working because non-linear effects
interfere heavily or because the system does not reach the aging regime.

Two examples of the agreement between dynamic and static data are shown 
in  figure {\ref{fig:sc_static}}.

\section{Conclusions}
We have studied the thermodynamics of a tiling model built by Wang tiles.
For this  kind of system we have found evidence of 
 a phase transition from a completely disordered
phase, in which the tiles on the plane are completely uncorrelated between
each other, to a phase in which they begin to present an organized, also if 
very complicated, structure. For $T\fd  0$ this structure becomes
an exactly matched tiling.

In order to characterize the phase of the system  we have determined
 an order parameter with a non-zero mean value below $T_c$. It is 
the overlap between two tilings built  
with the same boundary conditions on the lower and the left
 sides of the lattices.
 Our  data hint that it could
  have a non-trivial probability distribution under 
the phase transition.

Under the critical point the tiling system shows  the aging phenomenon:
the answer of the system to an external perturbation and the time
 autocorrelation function depend on
the history of the system.
This brings to a violation  of the fluctuation-dissipation theorem for $T<T_c$.
The integrated response function times the temperature ($S[C]$) shows
 a progressive
 bending when the system leaves the stationary regime until it reaches a
constant value ($X=0$)
 and  the dynamically and statically determined $q$ values
increase continuously with decreasing temperature.

 From this kind of behaviour it is not clear wether the model belongs to the
class of systems showing domain growth or it is rather  more similar to a 
spin glass in magnetic field. Indeed for very long times (small values of the
correlation function) the fluctuation-dissipation ratio goes eventually
to zero and it can not be excluded that the dynamics evolves through  domains
growth \cite{barrat}, 
even though in our case the nature of the domains of  tiles should still
be theoretically understood.
 Nevertheless for a very large interval of time, i.e. of  values
of $C$ in the plot of figure \ref{fig:sc_static}, the $S(C)$ is continuosly 
bending ($0<X<1$) like in a spin glass model \cite{mapa}\cite{pari}, especially
in the regions $0.3<C<0.6$, for $T=0.35$, and $0.4<C<0.7$, 
for $T=0.3$, as shown in figure \ref{fig:sc_static}.
For values of $C$ smaller than these the $S(C)$ curves flatten but the 
predictions obtained from the static behaviour (the $P(q)$ are shown in figure 
\ref{fig:pq}) are also nearly flat, making quite difficult the distinction
between a domain growth dynamics, where the response function is constant in 
the aging regime, and a more complicated behaviour
where even the response function shows aging.

Eventually we  show that our data are consistent with the equivalence between
the  equilibrium function $\int^{1}_{C} dq \ x(q)$ and
the dynamically determined  $S[C]$ as $t,t_w \fd \infty$.

{\bf {Acknowledgments}}
We warmly thank J.Kurchan for having stimulated  this work and for his advises.

\end{document}